\documentclass[journal]{IEEEtran}
\usepackage{amsmath,amsfonts}
\usepackage{mathtools}
\usepackage{hyperref}
\usepackage{array}
\usepackage[caption=false,font=normalsize,labelfont=sf,textfont=sf]{subfig}
\usepackage{textcomp}
\usepackage{stfloats}
\usepackage{xurl}
\usepackage{graphicx}
\usepackage{siunitx}
\usepackage{cite}
\usepackage{orcidlink}
\usepackage{svg}
\usepackage{graphicx}

\begin{document}

\title{High-performance computing enabled contingency analysis for modern power networks}

\author{Alexandre~Gràcia-Calvo\orcidlink{0009-0009-6728-6713},
        Francesca~Rossi\orcidlink{0000-0002-4791-151X},
        Eduardo~Iraola\orcidlink{0000-0002-0837-7086}, 
        Juan~Carlos~Olives-Camps\orcidlink{0000-0002-3122-1223}, \IEEEmembership{Member, IEEE,}
        and~Eduardo~Prieto-Araujo\orcidlink{0000-0003-4349-5923}, \IEEEmembership{Senior~Member, IEEE}
\thanks{The authors are with CITCEA-UPC and BSC, Barcelona, Spain. E-mail: alexandre.gracia@upc.edu, francesca.rossi@upc.edu, eduardo.iraola@bsc.edu, juan.carlos.olives@upc.edu, eduardo.prieto-araujo@upc.edu}
\thanks{Manuscript received November X, 2026; revised December X, 2026.}}

\markboth{IEEE Transactions on Power Systems,~Vol.~X, No.~X, November~2026}%
{Gr\`acia-Calvo \MakeLowercase{\textit{et al.}}: Critical Components in the IEEE 118-Bus Network}

\maketitle

\begin{abstract}
Modern power networks face increasing vulnerability to cascading failures due to high complexity and the growing penetration of intermittent resources, necessitating rigorous security assessment beyond the conventional $N-1$ criterion. Current approaches often struggle to achieve the computational tractability required for exhaustive $N-2$ contingency analysis integrated with complex stability evaluations like small-signal stability. Addressing this computational bottleneck and the limitations of deterministic screening, this paper presents a scalable methodology for the vulnerability assessment of modern power networks, integrating $N-2$ contingency analysis with small-signal stability evaluation. To prioritize critical components, we propose a probabilistic \textbf{Risk Index ($R_i$)} that weights the deterministic \textit{severity} of a contingency (including optimal power flow divergence, islanding, and oscillatory instability) by the \textit{failure frequency} of the involved elements based on reliability data. The proposed framework is implemented using High-Performance Computing (HPC) techniques through the PyCOMPSs parallel programming library, orchestrating optimal power flow simulations (VeraGrid) and small-signal analysis (STAMP) to enable the exhaustive exploration of massive contingency sets. The methodology is validated on the IEEE 118-bus test system, processing more than \num{57000} scenarios to identify components prone to triggering cascading failures. Results demonstrate that the risk-based approach effectively isolates critical assets that deterministic $N-1$ criteria often overlook. This work establishes a replicable and efficient workflow for probabilistic security assessment, suitable for large-scale networks and capable of supporting operator decision-making in near real-time environments.
\end{abstract}


\section{Introduction}

\IEEEPARstart {M}{odern} power networks are becoming increasingly complex, which heightens their vulnerability and necessitates rigorous assessment of their security posture \cite{Breton2025} against cascading failures initiated by simultaneous outages ($N-2$ contingencies). Conventional security analyses typically focus on single-outage scenarios ($N-1$) and often overlook the combined effects of multiple concurrent outages, which may have been the root cause of large-scale historical blackouts, such as the one that occurred on 28 April 2025 in Spain and Portugal~\cite{entsoe_28apr2025}. The increasing penetration of decentralized renewable energy sources and fast power electronics interfaced devices further complicates stability analysis \cite{stability_2023}, making system vulnerability assessment under severe multiple contingencies a critical task for grid operators.

The comprehensive assessment of $N-k$ contingencies has evolved primarily along three main axes: \textit{security assessment methodologies}, \textit{risk integration}, and \textit{computational acceleration}. Traditional deterministic contingency screening, while foundational, is often insufficient for modern complexity, prompting a shift toward     \textit{probabilistic security assessment (PSA)} which integrates component reliability data \cite{billinton1994reliability, ieee_rts96, agreira, perkin}. Previous works have successfully deployed Monte Carlo and probabilistic methods to rank risks based on power flow violations or transient stability criteria \cite{PSA_1_Pereira, shahzad, donde, contingency_screening_islanding}. However, these approaches often use simplified system models or fail to capture the full spectrum of post-contingency failure modes. A particularly persistent challenge is the integration of \textit{small-signal stability (SSS) analysis}, which is essential for detecting growing oscillatory modes, into exhaustive probabilistic frameworks \cite{bu, tabrizchi}. While tools for automated SSS modeling exist \cite{arevalo_sstoolbox}, they are computationally intensive. Furthermore, the explicit modeling of structural failure modes, such as \textit{system islanding}—a common precursor to cascading collapse—is frequently overlooked in large-scale screening methodologies. Addressing the computational load, High Performance Computing (HPC) has been successfully demonstrated to scale dynamic security assessments \cite{konstantelos2017}, though integrating this acceleration with open-source power flow platforms \cite{veragrid} and rigorous linearized EMT-based stability analyses for exhaustive $N-2$ evaluation remains a crucial implementation gap. Our work addresses these combined challenges:
\begin{enumerate}
    \item We move beyond $N-1$ and partial screening to provide an \textit{exhaustive $N-2$ analysis} of the test system.
    \item We define a \textit{multi-criteria risk index ($R_i$)}  that uniquely combines optimal power flow divergence, islanding, and, critically, small-signal instability.
    \item We demonstrate a replicable \textit{HPC parallelization strategy} using PyCOMPSs to make this complex, multi-criteria, and exhaustive analysis computationally tractable.
\end{enumerate}

This work presents a \textit{systematic and scalable methodology} for performing exhaustive $N-2$ contingency analysis, with the capacity to assess the impact on operational security through \textit{optimal power flow}, \textit{small-signal stability}, and \textit{islanding detection} (a multi-criteria scoring). The methodology is evaluated and validated on the IEEE 118-bus test system \cite{pena2017extended}, where results demonstrate the efficacy of the new probabilistic risk index compared to classic deterministic security criteria.
Specifically, we assess \textit{both single ($N-1$) and simultaneous double ($N-2$) outages} in AC lines, transformers and generators by \textit{modeling the post-contingency optimal power flow} and subsequently analyzing the eigenvalues of the linearized system state model to determine its stability. The ultimate goal is to define a robust and \textit{probabilistic ranking of the network's critical components}.

The contributions of this paper are:

\begin{itemize}
\item The definition and validation of a probabilistic \textit{multi-criteria Risk Index ($R_i$)}, which systematically combines AC optimal power flow feasibility, small-signal stability, and islanding criteria into a single risk metric weighted by component failure frequency.
\item Exhaustive processing of \num{57122} scenarios (encompassing single $N-1$ and ordered double $N-2$ outages), which serves as the foundational data set for the development and validation of the proposed probabilistic \textit{Risk Index} ($R_i$).
\item Implementation of a replicable \textit{HPC parallelization strategy} using the PyCOMPSs framework, ensuring that this complex, multi-criteria, and exhaustive analysis achieves scalability for potential execution within operational timeframes (below 15 minutes).
\end{itemize}

\par
The proposed methodology is tested and validated using the \textit{IEEE 118-bus test system}, a medium-sized yet highly interconnected benchmark network. This validation involves the exhaustive processing of \textit{\num{57122} contingency scenarios} (covering all $N-1$ and ordered $N-2$ outages), which represents a significant computational challenge. By successfully managing this massive scope, the framework demonstrates its effectiveness and scalability in identifying critical assets and providing a realistic risk ranking for modern power networks.


The paper is organized as follows. 
Section II provides a detailed description of the proposed \textit{Methodology}, including the exhaustive contingency analysis, the multi-criteria severity assessment (power flow, small-signal stability, and islanding), and the probabilistic Risk Index formulation. 
Section III details the \textit{HPC-Enabled Parallel Implementation} using PyCOMPSs. 
Section IV introduces the \textit{Case Study and Simulation Tools} (the IEEE 118-bus test system and specific reliability data). 
Section V presents the \textit{Results and Discussion}, including the contingency coverage, stability analysis, and the computed critical component ranking based on $R_i$, along with operational implications and future proposals. 
Finally, Section VI provides the \textit{Conclusion and Future Work}.

\section{Methodology}
\label{sec:method}

\subsection{Exhaustive Contingency Analysis}

The first step is to identify all the elements in the network that could fail, as well as all the possible combinations of two elements that could fail simultaneously. In particular, we consider contingencies involving transmission lines, power transformers and generators. The following procedure is implemented to determine under what contingencies the system cannot be operated:

\begin{enumerate}
    \item Start from the base state of the system, compute the pre-contingency optimal power flow, and verify that the system is initially stable to establish a reference.
    
    \item For each element $i$ (line, transformer or generator) in the network:
    \begin{enumerate}
        \item \textbf{$N-1$ contingency (first outage):} Simulate the outage of element $i$, compute the optimal power flow, perform small-signal stability analysis, check for island formation, and store the results.

        \item For each remaining element $j$ such that $j \ne i$:
        \begin{enumerate}
            \item \textbf{$N-2$ contingency (second outage):} With $i$ already out, simulate the outage of $j$. Compute optimal power flow, assess stability, check for islands and store results.
            \item Restore element $j$ to service.
        \end{enumerate}

        \item Restore element $i$ to service.
    \end{enumerate}
    
    \item Analyze the complete set of results to identify the most critical elements.
\end{enumerate}

As indicated in the algorithm, during each contingency simulation we check for possible separation into electrical islands. In general, a division of the system into multiple areas can affect the analysis: often, only the main island interconnected to the largest generation should be taken as reference. In this initial stage, any scenario that results in islanding is marked as critical and kept in the study for specific revision afterwards.

\subsection{Multi-Criteria Severity Assessment}

\label{sec:severity}
The severity, $S_c \in \{0, 1\}$, of a post-contingency state $c$ is determined by the consideration of distinct criteria. For the contingency to be classified as severe ($S_c=1$), at least one of the following conditions must be fulfilled:

\subsubsection{Optimal Power Flow Feasibility and Operational Limits}
A contingency $c$ is considered severe if the AC optimal power flow solution does not converge, suggesting a structurally non-feasible operating point, or if the converged solution violates pre-defined operational limits (e.g., thermal limits, voltage bounds).

\subsubsection{Small-Signal Stability Evaluation}

To perform the small-signal analysis, we construct and linearize the complete system model at the operating point obtained from the optimal power flow solution. The general state-space formulation for small perturbations is applied, following established control theory principles for power systems \cite{pal2006}.


Small-signal stability is assessed from the eigenvalues of the state matrix, and any contingency leading to at least one eigenvalue with non-negative real part ($Ri(\lambda_i) \ge 0$) is classified as unstable.


\subsubsection{System Integrity: Islanding Detection}
\label{sec:illes}

A key component in determining the severity of a contingency ($S_c$) is the structural integrity of the network. Any contingency, whether $N-1$ or $N-2$, that causes the fragmentation of the system into multiple electrical islands is considered a severe failure event. These islands may suffer from fatal generation-load imbalances or the loss of the slack bus, potentially leading to cascading collapse.

To evaluate this phenomenon, our methodology incorporates a topological connectivity analysis after each contingency simulation, based on graph-theoretic principles, in line with graph-based frameworks for vulnerability assessment in power systems \cite{sperstad_vulnerability_graph, veragrid}.

The procedure is as follows:
\begin{enumerate}
    \item After simulating the disconnection of one or two elements, the numerical model of the network in its post-contingency state is generated.
    \item A graph-analysis algorithm (similar to a Depth-First Search) is applied to the resulting topology to identify all connected subgraphs.
    \item The number of electrically isolated subnetworks (islands) is then counted.
\end{enumerate}

If the number of resulting islands is greater than one, the scenario is automatically classified as a failure of maximum severity ($S_c = 1$), and its associated risk ($R_i$) is computed accordingly. This ensures that any loss of network integrity is appropriately penalised in the final risk index.

\par
\textit{Note on the applicability to $N-1$ analysis:} Although the primary focus of this work is the exhaustive $N-2$ analysis, the proposed multi-criteria severity assessment and probabilistic framework remain highly valuable for single-outage ($N-1$) scenarios. Traditional $N-1$ security analysis often relies solely on steady-state limit violations. By integrating Optimal Power Flow (OPF) feasibility, structural islanding detection, and, critically, \textit{small-signal stability (SSS)} evaluation into the severity score ($S_c$), the methodology provides a far more rigorous safety assessment for $N-1$ events. Furthermore, by utilizing component failure frequencies ($\lambda_i$), the resulting $R_i$ index elevates the $N-1$ evaluation from a purely deterministic screening (where all severe failures are equally prioritized) to a \textit{probabilistic risk ranking}, allowing operators to focus on the single component failures that contribute the most to the expected annual system risk.

\subsection{Probabilistic Risk Index Formulation}

To move from a purely deterministic analysis (where all failures are considered equally likely) to a probabilistic perspective, we weigh the severity of each contingency by its frequency of occurrence. The key concept is that
\[
\textbf{Risk} = \text{Frequency} \times \text{Severity}.
\]

For each network component $i$ (line, transformer, or generator), one essential reliability parameter is required: the \textit{failure rate} ($\lambda_i$), defined as the expected frequency of failures of component $i$ (in failures per year) and related to the mean time to failure (MTTF) by $\lambda_i = 1 / \text{MTTF}_i$.

We define the expected annual frequency for each contingency scenario:

\begin{enumerate}
    \item \textbf{$N-1$ frequency (failure of $i$):} the failure rate of the component,
    \begin{equation}
        F_i = \lambda_i.
    \end{equation}

    \item \textbf{$N-2$ frequency (failure of $i$ and $j$):} the average rate at which $j$ fails while $i$ is already out of service (assuming independence),
    \begin{equation}
        F_{i,j} = \lambda_i \lambda_j.
    \end{equation}
\end{enumerate}

It is essential to note that the frequency of an $N-1$ failure (e.g., $10^{-1}$ events/year) is several orders of magnitude higher than that of an $N-2$ failure (e.g., $10^{-5}$ events/year). This natural difference in frequency replaces the need for artificial weighting to prioritize $N-1$ failures.

The other important term for characterizing risk is severity. In this work, we define the severity of the $i$-th element as a binary variable whose value is obtained in the deterministic analysis. Specifically, $S_i = 1$ if the contingency of element $i$ leads to a failure (non-convergence of optimal power flow, small signal instability, or undesired island formation), and $S_i = 0$ otherwise. When analyzing the failure of a second element $j$, the severity is denoted as $S_{i,j}$, but the value is obtained in an analogous manner.




The Risk Index for a component $i$ $R_i$ is computed as the sum of all risk contributions (frequency $\times$ severity) from every contingency scenario in which $i$ participates, as follows:

\begin{equation}
R_i = F_i S_i + \sum_{j \neq i} F_{i,j} S_{i,j},
\label{eq:Re}
\end{equation}
where:
\begin{itemize}
    \item $R_i$ is the risk index for component $i$ (failure-events/year),
    \item $F_i$ is the $N-1$ failure frequency of $i$ (i.e., $\lambda_i$),
    \item $S_i$ is the severity of the $N-1$ failure of $i$ (1 if failure, 0 otherwise),
    \item $j$ denotes another component in the network ($j \neq i$),
    \item $F_{i,j}$ is the $N-2$ failure frequency of the pair $(i,j)$ (i.e., $\lambda_i \lambda_j$),
    \item $S_{i,j}$ is the severity of the $N-2$ failure of $(i,j)$.
\end{itemize}

\par
It is important to note that \eqref{eq:Re} 
uses a \textit{sum} to aggregate risk. The index $R_i$ does not represent the risk of a single scenario, but rather the \textit{total risk contribution} of component $i$ to the system. It is therefore computed as the sum of the risk of the $N-1$ scenario (failure of $i$ alone) plus the sum of all individual risks of the $N-2$ scenarios in which $i$ is involved (failure of $i$ together with any other component $j$). The resulting $R_i$ has units of failure-events/year, representing the expected frequency of systemic failures in which component $i$ is one of the contributors.

This index $R_i$ captures the expected contribution of each component to the total number of systemic failures per year. A component with a high $R_i$ is therefore more critical for the secure operation of the grid, either because it fails frequently (high $F_i$) or because its failure (alone or in combination) is particularly destabilising (high $\sum S_{i,j}$).

\section{HPC-Enabled Parallel Implementation}

Due to the large number of cases that had to be performed in the $N-2$ contingency analysis, the entire process has been parallelized using the PyCOMPSs framework from BSC \cite{nord4_bsc}. PyCOMPSs \cite{pycompss_softx, pycompss_tejedor2017} is a Python-based programming environment that facilitates the definition of parallel tasks and their execution on distributed environments (HPC clusters, cloud). In our case, each contingency analysis (optimal power flow + stability analysis + island detection) is encapsulated as an independent task (typically requiring around 30 seconds depending on hardware conditions). Execution was performed on the Nord4 cluster at the Barcelona Supercomputing Center, allowing each contingency to be assigned a dedicated subset of cores for a short period in a burst of computational power.

This approach drastically reduced the total computation time: as resources are increased, the process scales almost linearly, in contrast to the exponential growth in the number of contingencies as their order increases. In parallel, we maintained a JSON file to store the results of each simulation (stability outcome, power flow and optimal power flow convergence, type of contingency—single or double—elements involved and whether islands were formed, etc.) without introducing I/O bottlenecks.

The general procedure is as follows:
\begin{enumerate}
    \item Python functions that compute optimal power flows, build the stability model for each contingency, and detect islanding are written sequentially and decorated as PyCOMPSs tasks.
    \item The PyCOMPSs runtime builds a dependency graph and schedules tasks across the available CPU cores.
    \item After computation is completed, the results are collected and aggregated (convergence, stability, island detection, etc.).
\end{enumerate}

This approach enables horizontal scaling of the exhaustive analysis to networks of the medium size and, potentially, larger systems, and lays the foundations for integrating a real-time decision-support assistant into the system operation center.

\section{Case Study and Simulation Tools}
\label{test}
\begin{figure}[hbt!]
    \centering
    \includegraphics[width=0.45\textwidth]{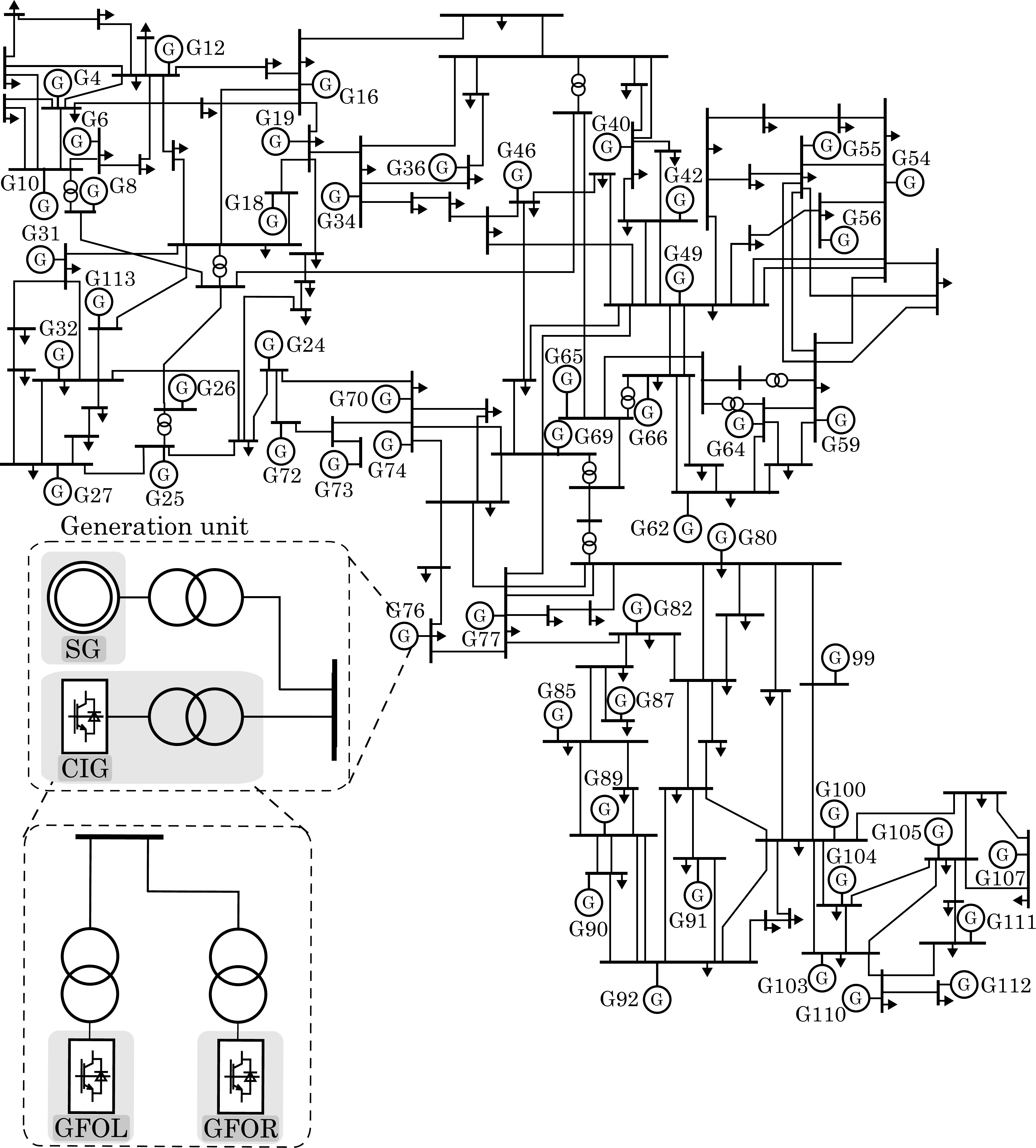}
    
    \caption{Topological representation of the IEEE 118-bus test system.}
    \label{fig:ieee118_topology}
\end{figure}

This section presents the results of evaluating the proposed methodology on the IEEE 118-bus network\cite{pena2017extended}, a classic benchmark system for power system stability and vulnerability assessment.
The network consists of 118 high-voltage buses interconnected by 175 transmission lines and 53 generators or converters, and 11 transformers. Figure \ref{fig:ieee118_topology} presents a single-line diagram of the network. For optimal power flow simulation and contingency analysis we use VeraGrid \cite{veragrid}, an open Python environment that includes tools for AC optimal power flow, OPF, and system dynamics. The base configuration is obtained through an AC optimal power flow, and island detection is performed using the VeraGrid framework. For small-signal stability analysis in EMT \cite{arevalo_sstoolbox}(STAMP Tool), we construct state-space linearized models of the entire system after each single and double contingency, incorporating generator dynamic equations and voltage regulation controls.

Since the IEEE 118-bus test case is a benchmark system lacking specific historical outage data, representative reliability parameters have been adopted based on standard values found in the literature \cite{billinton1994reliability}, with generator failure rates specifically aligned with the IEEE Reliability Test System (RTS-96) \cite{ieee_rts96}. These default values, summarized in Table~\ref{tab:reliability}, are used to demonstrate the proposed methodology assuming typical failure characteristics for high-voltage components. It should be noted that, while uniform rates are applied here for this benchmark study, the proposed framework fully supports component-specific failure rates enabling the integration of real operational data.

\begin{table}[htbp]
    \caption{Assigned Reliability Parameters}
    \label{tab:reliability}
    \centering
    \begin{tabular}{l c c}
    \hline
    \textbf{Component} & \textbf{Failure rate ($\lambda_i$)} & \textbf{MTTF} \\
    \hline
        Lines & $0.05 \text{ yr}^{-1}$ & 20 years \\
        Transformers & $0.02 \text{ yr}^{-1}$ & 50 years \\
        Generators & $0.10 \text{ yr}^{-1}$ & 10 years \\
    \hline
    \end{tabular}
\end{table}

\section{Results and Discussion}
\label{results}

A total of \num{57122} contingency scenarios were evaluated, covering all $N-1$ and $N-2$ combinations considered in the study. Of these, \num{51493} cases were classified as stable (or secure) according to the adopted criteria (optimal power flow convergence, absence of island formation and all eigenvalues with negative real part), which represents approximately \qty{90.15}{\percent} of all analyzed scenarios. The remaining \num{5629} cases were identified as unstable (around \qty{9.85}{\percent}), either due to divergence in the optimal power flow, the formation of electrical islands or small-signal instability detected through eigenvalue analysis.

These unstable situations correspond to contingency conditions in which the system loses operational security. The following subsections examine the distribution of these failures, the most relevant patterns observed and their implications for the computation of the risk index.

All \num{57122} contingency cases were processed on the Nord4 HPC cluster \cite{nord4_bsc}. We used 8 compute nodes, each with 48 CPU cores and 96 GB of main memory. Because the workflow consists of independent contingency evaluations, it parallelised efficiently across the cluster. The full analysis finished in about five hours of wall-clock time, indicating that the proposed method is computationally tractable at large scale.

It is worth noting that the method is \textit{embarrassingly parallel}, and therefore its execution time can be reduced almost linearly by allocating additional computational resources. This aligns with the operator-support use cases discussed in the conclusions, where time budgets on the order of 15 minutes may be required.

\subsection{Contingency Coverage and Stability}

A total of \num{57122} $N-1$ and $N-2$ contingency scenarios were analysed, including symmetric inverted combinations to ensure full coverage. The vast majority of these scenarios remain stable, except for approximately 1390 cases that resulted in numerical errors or non-convergent solutions. Most of these problematic cases appear to be associated with situations where the slack bus becomes isolated or located inside an island, which suggests that an alternative slack assignment could improve convergence reliability.

Among all $N-2$ simulations, a large proportion (\qty{90.15}{\percent}) converge successfully and show all eigenvalues with negative real parts, indicating small-signal stable post-contingency operation. By contrast, \qty{9.85}{\percent}  of scenarios exhibit at least one eigenvalue with positive real part, revealing small-signal instability or lead to the formation of one or more electrical islands, which we classify as severe events since they can lead to loss of reference, imbalance between load and generation, or cascading collapse.

These critical cases represent the most relevant patterns from the operational perspective, and they often correspond to specific structural weaknesses in the network topology.

The computation of the risk index $R_i$ allows us to quantify how much each network component contributes to the overall likelihood of system failure. This provides a powerful tool for prioritizing maintenance, designing protection schemes, and preparing contingency strategies.

Overall, the IEEE 118 system shows a robustness level of approximately \qty{90.15}{\percent} under $N-2$ analysis. However, the identified critical scenarios highlight that certain combinations of failures pose a significantly higher threat to operational security. These results offer a practical path for grid operators, who could focus their preventive actions on the small subset of components that dominate the system-level risk.

\subsection{Interpretation of the Risk Index Values}

The computed Risk Index ($R_i$) provides a quantitative measure of the expected contribution of each component to system-level failure events per year. Its interpretation is closely tied to both the reliability parameters (failure rates $\lambda_i$) and the observed severity of the simulated contingencies ($S_c$). Understanding the order of magnitude of $R_i$ is essential for assessing the operational relevance of the results.

A value of $R_i$ close to zero indicates that the component rarely participates in severe scenarios. This may occur for two distinct reasons: either the component has a low failure rate (large MTTF) or its individual outage and the combinations in which it participates do not lead to instability, island formation or loss of optimal power flow convergence. In practical terms, components with $R_i \approx 0$ have negligible impact on global security, even if they are part of the network topology.

On the other hand, values of the order of $R_i \approx 1.4$ represent a dramatically different situation. For example, a line with $R_i = 1.472$ contributes on average to almost one and a half severe system failures per year when accounting for both its $N-1$ and $N-2$ interactions. Since the failure rate of transmission lines is relatively low ($\lambda_i = 0.05$ year$^{-1}$), such a high value of $R_i$ can only emerge when the component is involved in a very large number of $N-2$ contingencies with $S_{i,j} = 1$. This means that its structural position in the network makes it systematically part of combinations that reduce damping, induce instability or generate islands.

In this study, the resulting distribution of $R_i$ spans approximately two orders of magnitude. The smallest values are below $10^{-3}$ failure events per year, typical of components that neither fail often nor lead to severe outcomes when they do. Intermediate values, in the range $10^{-2}$ to $10^{-1}$, correspond mostly to generators and a subset of transformers with moderate participation in unstable combinations. Finally, the highest values, between $0.5$ and almost $1.5$ failure events per year, are dominated by transmission lines that appear repeatedly in severe $N-2$ combinations and whose reliability parameters amplify the accumulated risk.

These ranges illustrate that system-level vulnerability is not evenly distributed across the network. Instead, a small set of components produces most of the expected annual risk, while the majority contribute negligibly. This asymmetry is critical for prioritising maintenance, targeted monitoring and protection strategies.

\subsection{Critical Components According to the Risk Index}

\begin{figure}[!t]
    \centering
        \includegraphics[width=\linewidth]{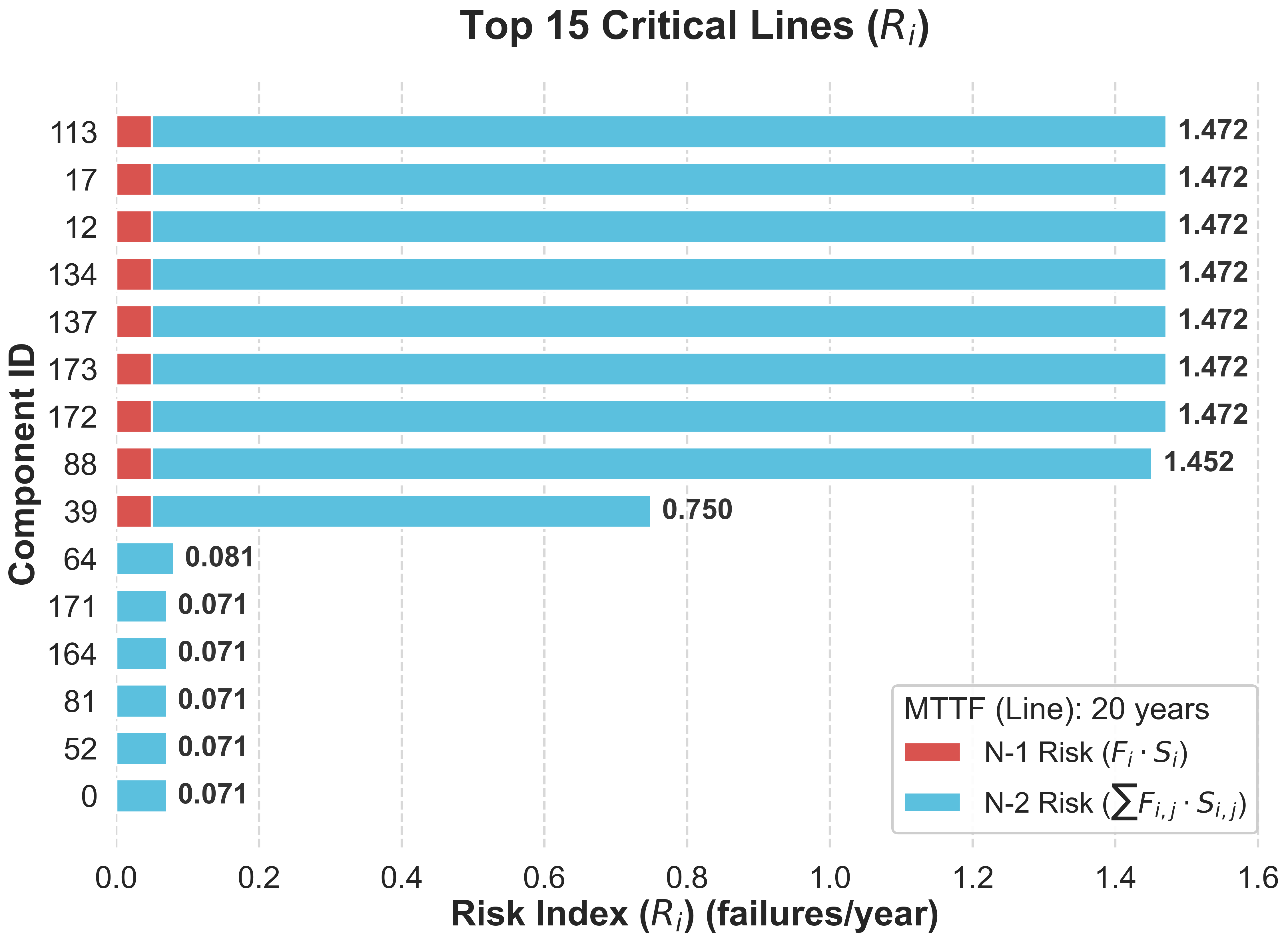}
    \caption{Risk Index ($R_i$) for transmission lines.}
    \label{risk_index_for_lines}
\end{figure}

\begin{figure}[!t]
    \centering
        \includegraphics[width=\linewidth]{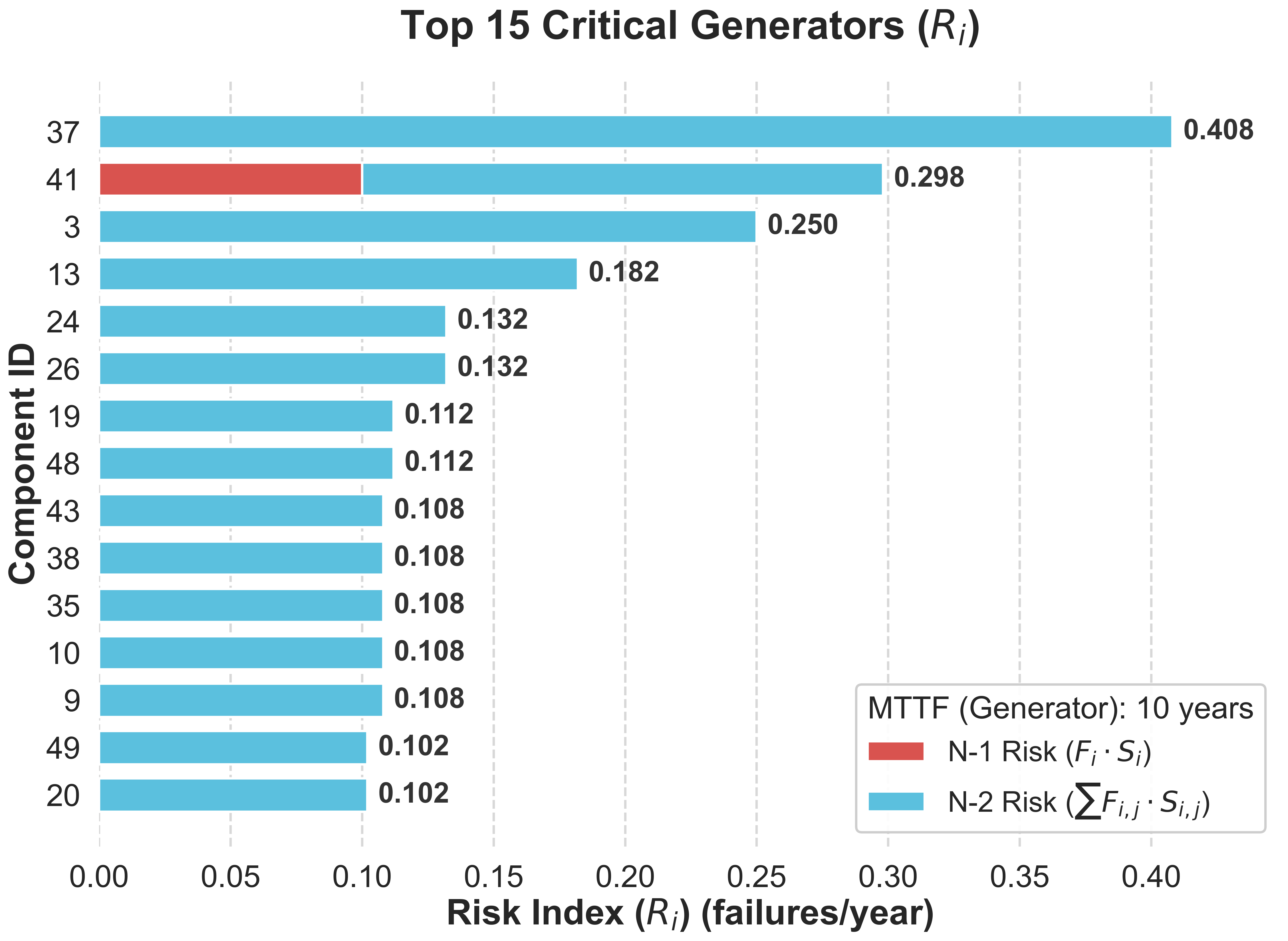}
    \caption{Risk Index ($R_i$) for generators.}
    \label{risk_index_for_generators}
\end{figure}

\begin{figure}[!t]
    \centering
        \includegraphics[width=\linewidth]{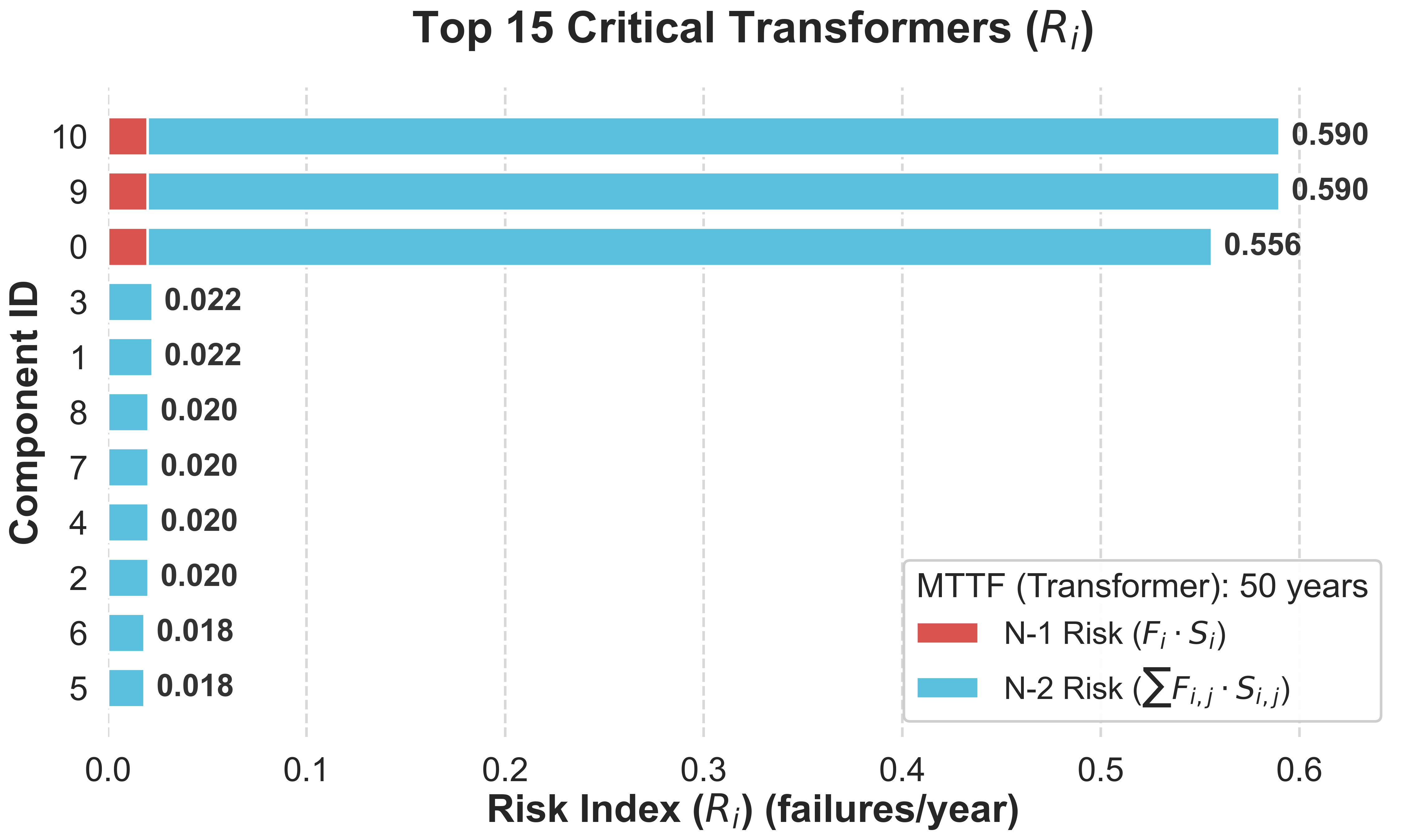}
    \caption{Risk Index ($R_i$) for transformers.}
    \label{risk_index_for_transformers}
\end{figure}

\begin{figure}[!t]
    \centering
        \includegraphics[width=\linewidth]{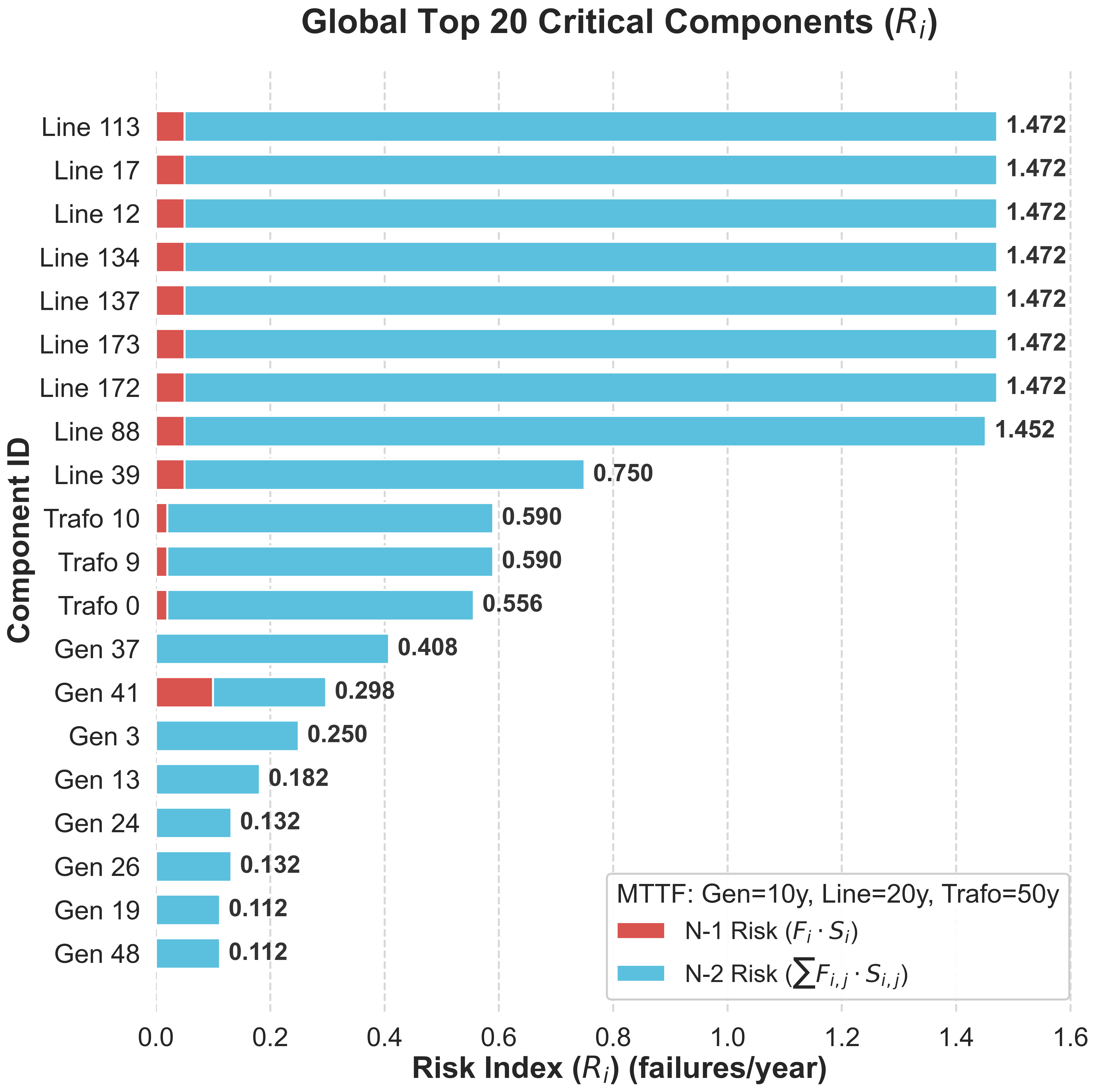}
    \caption{Combined Risk Index ($R_i$) for all elements.}
    \label{risk_index_for_combined}
\end{figure}

Figures \ref{risk_index_for_lines}, \ref{risk_index_for_generators}, and \ref{risk_index_for_transformers} show the risk index $R_i$ for lines, generators and transformers respectively. The combined ranking across all elements is presented in Fig. \ref{risk_index_for_combined}, which consolidates the global top 20 most critical components.

A clear pattern emerges in the combined ranking. The first seven positions correspond to transmission lines (IDs 113, 17, 12, 134, 137, 173 and 172), all with nearly identical values around $R_i \approx \num{1.47}$ failure events per year, followed closely by Line 88 ($R_i \approx \num{1.45}$). This similarity is not coincidental. It results directly from the uniform MTTF assigned to all transmission lines (20 years, corresponding to a failure rate $\lambda_i = 0.05$ year$^{-1}$). Since these lines also participate very frequently in severe $N-2$ combinations, the resulting products $\lambda_i \lambda_j$ accumulate to extremely similar $R_i$ values. In these cases, the blue bars dominate, showing that the overwhelming majority of their risk contribution comes from $N-2$ interactions rather than from their individual $N-1$ outage.

Below this top cluster, a second tier appears. Line 39 has $R_i \approx 0.75$, followed by a group of three transformers (IDs 10, 9 and 0) with values between $0.56$ and $0.59$. Again, this pattern strongly reflects the reliability assumptions. Transformers were assigned an MTTF of 50 years ($\lambda_i = 0.02$ year$^{-1}$), which yields lower $N-1$ risk and reduced $N-2$ frequency. The resulting sharp drop between transmission lines and transformers illustrates how reliability parameters directly influence the probabilistic ranking.

Generators appear mostly in the lower half of the distribution. Generator 37 exhibits the highest generator risk ($R_i \approx \num{0.41}$), almost entirely due to $N-2$ contributions. Generator 41 ($R_i \approx \num{0.30}$) displays a significant red bar, showing that even its individual $N-1$ failure is classified as severe. The remaining generators (IDs 3, 13, 24, 26, 19 and 48) show much smaller $R_i$ values, reflecting both their lower frequency of participation in unstable $N-2$ scenarios and their higher assigned failure rate ($\lambda_i = 0.10$ year$^{-1}$, MTTF 10 years). This last point demonstrates an interesting phenomenon: despite generators having a higher intrinsic failure rate, they contribute far less to systemic instability than transmission lines. This means that, in this particular system, structural vulnerability is concentrated in the transmission network rather than in generation assets.

It is important to note that although we use category-wide default MTTF values for lines, transformers and generators, the implementation allows assigning individual MTTFs to specific components. This flexibility in the proposed methodology makes the risk index directly applicable to real systems in which reliability is heterogeneous and asset specific. By modifying element-level MTTF values, operators can immediately observe how the risk profile shifts as a result of ageing infrastructure, maintenance strategy or external environmental stressors.

The results demonstrate that system-level risk is highly concentrated in a small subset of components. Most of the expected annual failure events originate from combinations involving transmission lines, while transformers and generators play a secondary role. This highlights the importance of double-outage analysis. Many of the dominant contributions arise from $N-2$ severity rather than $N-1$ failures, which reinforces the need to evaluate interactions that are entirely invisible under a classical $N-1$ operational criterion.

In practical terms, the risk index $R_i$ helps operators distinguish between components that are statistically more likely to fail and those whose failures, even if rare, have disproportionate consequences. It provides a unified and quantitative framework to support asset prioritisation, maintenance planning and real-time contingency assessment.

Although this study focuses on $N-1$ and $N-2$ contingencies, real systems may experience more complex outage patterns. Nonetheless, $N-2$ contingency analysis combined with small-signal stability assessment offers a solid foundation for understanding real-world vulnerabilities. The structured methodology allows scalable analysis and future integration into operational practice, which is particularly valuable in contingency management.

This study demonstrates the feasibility of systematic $N-2$ analysis with small-signal stability evaluation on a medium-sized system. However, in larger networks the computational workload increases substantially and additional phenomena may arise.

A first-order probabilistic model has already been incorporated, using failure rates ($\lambda_i$) to weight contingency severity and compute the Risk Index ($R_i$). This risk-based (frequency $\times$ severity) perspective is more realistic than a purely deterministic ranking.

Several improvements could be explored:

\begin{itemize}
    \item \textbf{Common-Cause Failures:} The current model assumes $N-2$ events are independent. In reality, storms or human errors may cause simultaneous failures (e.g., two lines on the same tower). Incorporating such probabilities would yield a more accurate $N-2$ risk estimate.
    \item \textbf{Non-binary Severity:} Severity is currently binary ($S_c \in \{0,1\}$). A more refined model could treat severity as continuous, e.g., load shed in MW, voltage depression indices, or damping ratios, allowing differentiation between mild and catastrophic failures.
    \item \textbf{Uncertainty Analysis:} Failure rates $\lambda_i$ are uncertain. A Monte-Carlo analysis on $\lambda$ would enable confidence intervals for the risk ranking instead of a single point estimate.
    \item \textbf{Component-Specific Reliability:} The developed framework supports the assignment of specific MTTF values to each individual network element. Although this study applied uniform failure rates per asset class (lines, transformers, generators) for standardization purposes within the benchmark, the methodology is fully capable of processing heterogeneous reliability data. This allows for a more precise assessment in real-world scenarios where components exhibit varying failure probabilities based on their specific age, condition, or maintenance history.

\end{itemize}

In parallel, computational efficiency could be improved using machine-learning methods or heuristics to reduce the search space for preliminary screening of weak areas to filter out minor contingencies before running a full powerflow AC simulation. Further automation of islanding analysis (e.g., via reconnection heuristics or graph-theoretic tools such as improved Dijkstra variants) would also be beneficial.

\subsection{Operational Implications of N-1 and N-2 Failures}

From an operational perspective, the distinction between $N-1$ and $N-2$ contingencies is essential for understanding how the system should react to unexpected outages and how operators can anticipate cascading failures.

An $N-1$ failure corresponds to the outage of a single element, such as a transmission line, a transformer or a generator. Modern transmission systems are typically designed to withstand any $N-1$ event without losing operational security, meaning that voltage limits must remain acceptable, flows must continue to satisfy thermal limits and the system must preserve small-signal stability. When an $N-1$ contingency occurs, operators rely on predefined corrective actions that are well established in operational procedures. These actions may include generation rescheduling, activation of reserves, topology reconfiguration, re-dispatch of power flows through alternative paths or adjustments in voltage and reactive support devices. In most systems, these actions can be performed within a few minutes and do not require emergency procedures. Therefore, an $N-1$ failure is expected to be a manageable event that does not compromise overall system integrity.

In contrast, an $N-2$ contingency represents the simultaneous outage of two components. This type of event is much more challenging to handle because the consequences are harder to predict and the system may not be designed to tolerate such conditions under all operating states. An $N-2$ failure may lead to severe overloads on neighbouring elements, rapid voltage deterioration, loss of synchronism or the creation of electrically isolated islands. In such cases, corrective actions available to the operator become more restrictive. Actions that are feasible after an $N-1$ failure, such as rerouting flows or re-dispatching generation, may no longer be sufficient, especially if both outages affect major corridors or essential components. Emergency measures, such as controlled load shedding, temporary isolation of vulnerable areas or rapid reserve activation, may be necessary to stabilise the system.

The key difficulty is that $N-2$ failures often occur before operators have enough time to react to the initial $N-1$ event. If the system is already weakened by the first outage, the second outage can push it beyond its stability margin, leaving little time for corrective action. This highlights the importance of identifying the specific $N-2$ combinations that lead to instability or island formation, as these combinations reveal structural weaknesses that are hidden under classical $N-1$ analysis.

In summary, $N-1$ failures represent expected and generally manageable operational disturbances, while $N-2$ failures expose the system to high-impact situations that may require emergency actions and can escalate into cascading events. Understanding how each component contributes to these two classes of failures enables operators to prioritize preventive strategies, reinforce vulnerable areas and prepare focused response plans for the most critical multi-outage scenarios.

\subsection{Real-Time Operator Support Tool}

To bring the results of this analysis into operational practice, we propose an interactive operator support tool. Its goal is to integrate automated contingency calculations and risk indices ($R_i$) into a practical decision-support environment. In near real-time, the tool would highlight the most vulnerable components and suggest preventive or corrective actions. A simple graphical interface could include a network topology visualization, color-coded by element criticality (lines, transformers, generators), along with alarms or notifications when high-risk situations arise. This would facilitate rapid situational awareness and informed decision-making.

The tool should be structured into several modules:

\begin{itemize}
    \item \textbf{Real-time data management:} Interfaces with operational data sources (SCADA) to obtain the current grid state (flows, generation, voltages, topology). It continuously updates internal models without interfering with ongoing operation.

    \item \textbf{Contingency simulation engine:} Integrates VeraGrid and HPC infrastructure to execute power-flow and small-signal stability calculations quickly. It could run in the background or on demand, automatically simulating the most relevant $N-2$ scenarios or enabling operator-driven what-if analysis.

    \item \textbf{Risk index computation and visualization:} Processes simulation outputs to compute the criticality index $R_i$ for each element in near real-time. It highlights recurring unstable components and their associated severity, marking them as “hot spots”. This information can guide operational actions (e.g., protective relays adjustment, control tuning, load reallocation).

    \item \textbf{Recommendation and visualization panel:} Provides an intuitive interface displaying network maps or graphs with risk indicators, lists of critical contingencies and suggested actions. For example, the most critical line or generator could be highlighted, with messages such as “reduce output”, “activate reserves” or “redirect flows” based on predefined rules or prior analysis. It may also show time-evolution trends of critical eigenvalues or global $R_i$ values.
\end{itemize}

A modular architecture could integrate these components through APIs or Python libraries with VeraGrid, along with a database for simulation logs and configurations. The essential goal is collaborative operation: giving operators near real-time guidance on the most dangerous multi-failure scenarios and suggesting preventive measures.

This bridges the gap between offline analysis and operational tools, turning the methodology into actionable decision-support.

\section{Conclusion and Future Work}
\label{conclusion}

We have presented a structured methodology for identifying critical components in the IEEE-118 system through exhaustive $N-2$ contingency enumeration, small-signal stability analysis and islanding detection. Using VeraGrid and Python, we modelled the impact of simultaneous outages on AC optimal power flow and subsequently constructed linearized EMT-based state-space models to assess dynamic stability. The analysis reveals that a non-negligible fraction of the \num{57122} $N-1$ and $N-2$ combinations leads to instability or island formation, exposing specific components as systematic drivers of worst-case scenarios.

The proposed Risk Index ($R_i$) integrates deterministic severity and probabilistic failure frequency into a unified metric, enabling a meaningful ranking of components according to their expected contribution to system-level risk. Results show that risk is highly concentrated: a small subset of transmission lines dominates the global risk due to both their structural position in the network and the assumed reliability parameters (MTTF), while only a few transformers and generators make significant contributions. This aligns with the flexibility of the method, which allows individual MTTF assignments and can therefore adapt to real asset conditions in operational networks.

This probabilistic-deterministic framework offers a more realistic perspective than traditional $N-1$ security analysis, highlighting the importance of $N-2$ interactions that are otherwise invisible to classical operational criteria. The approach demonstrates strong potential to evolve into a proactive grid-security tool, capable of anticipating vulnerabilities and guiding targeted interventions. Future improvements include integrating real-world reliability data, exploring higher-order contingencies ($N-k$ with $k>2$), incorporating non-binary severity metrics and extending the methodology to larger or time-varying network conditions.

A particularly relevant direction is the development of a near real-time operator support system that embeds contingency simulation, risk index computation and graphical visualization. Thanks to HPC scalability and parallelisation, the full computational workflow could be executed within operational timeframes $(t<15\ \text{minutes})$, enabling informed decision-making and early warning mechanisms in control-room environments.

Overall, the proposed methodology not only deepens the theoretical understanding of multi-contingency security in networks such as IEEE-118 but also lays the foundation for practical tools with direct applicability in industrial and research settings. It creates a bridge between offline probabilistic risk assessment and actionable real-time operational support, while supporting a systematic, reproducible, quantitative way to identify vulnerabilities, prioritize mitigations, support operators, justify planning decisions, and explore the effects of failures.

\section*{Acknowledgment}
\label{acknowledgement}

The Project TED2021-130351B-C21 (HP2C-DT) is funded by MICIU/AEI/10.13039/501100011033 and by the European Union NextGenerationEU/PRTR.

\bibliographystyle{IEEEtran}

\vfill

\end{document}